\documentclass[12pt]{article}

\usepackage{amssymb}

\usepackage{latexsym}

\newcommand{\bea}{\begin{eqnarray}}

\newcommand{\be}{\begin{equation}}

\newcommand{\eea}{\end{eqnarray}}

\newcommand{\ee}{\end{equation}}

\def\le{\left}

\def\ri{\right}

\def\part{\partial}

\def\a{\alpha}

\def\b{\beta}

\def\e{\epsilon}

\def\m{\mu}

\def\n{\nu}

\def\th{\theta}

\def\s{\sigma}

\def\t{\tau}

\def\O{\Omega}

\begin{document}


\title{REDUCTION OF THE GEOMETRICAL MASS OF THE
REISSNER-NORDSTR\"OM SPACETIME}

\author{C. Barbachoux$^1$\thanks{e-mail: barba@ccr.jussieu.fr},
J. Gariel$^1$\thanks{e-mail: gariel@ccr.jussieu.fr}, G.
Marcilhacy$^1$ and N. O. Santos$^{1,2,3}$
\thanks{e-mail: santos@ccr.jussieu.fr and nos@cbpf.br}
\\ \\
{\small $^1$LRM-CNRS/UMR 8540, Universit\'e Pierre et Marie Curie,
ERGA,}\\
{\small Bo\^{\i}te 142, 4 place Jussieu, 75005 Paris Cedex 05, France.}\\
{\small $^2$Laborat\'orio Nacional de Computa\c{c}\~{a}o
Cient\'{\i}fica,}\\
{\small 25651-070 Petr\'opolis RJ, Brazil.}\\
{\small $^3$Centro Brasileiro de Pesquisas F\'{\i}sicas,}\\
{\small 22290-180 Rio de Janeiro RJ, Brazil.}}
\maketitle

\begin{abstract}
We derive the Teixeira, Wolk and Som method \cite{TWS}, for obtaining
electrostatic
solutions from given vacuum solutions, in its inverse form. Then we use
it to obtain the
geometrical mass $M_S$ in the Schwarzschild spacetime, and we find
$M_S^2=M^2-Q^2$, where $M$ and $Q$ are, respectively, the mass and charge
parameters of the Reissner-Nordstr\"om spacetime. We compare $M_S$ to the
corresponding active gravitational mass and mass function.

\end{abstract}
\pagebreak

\section{Introduction}
For a bounded spherically symmetric static distribution of matter the
total mass is well defined. We know that outside of the
distribution the spacetime must be described by the Schwarzschild
spacetime, therefore it follows from the junction conditions \cite{Misner}
that the total mass of the system is equal to the Schwarzschild mass
parameter \cite{Rosen}. However the definition of the total mass
content within a given spherical surface inside a non vacuum spacetime is
not unique. This ambiguity has been the subject of long discussions, giving
rise to different definitions of energy (see \cite{Herrera} and references
therein).

Our aim in this article is to revisit the question of mass in the
Reissner-Nordstr\"om (RN) spacetime.

The study of charged bodies in Einstein's theory contributes to a better
understanding of the structure of spacetime, as it is showed by many new
solutions recently studied for electrovacuum (see \cite{Miguelote,Socorro}
and references therein) and different charged sources
(see \cite{Guilfoyle,Ivanov} and references therein).
For the spherically symmetric static spacetime the solution
of the coupled Einstein and Maxwell equations is the RN
solution. This solution is the unique black hole solution with a regular
event horizon and assymptotically flat behaviour. The RN spacetime
provides a more general framework to study the structure of Schwarzschild
spacetime. The fact that RN solution has two horizons, an external
event horizon and an internal Cauchy horizon, provides a convenient bridge
to the study of the Kerr solution, as pointed out by Chandrasekhar
\cite{Chandra}. Furthermore, the RN field can be used as a simple model
of the electron, as suggested by Bonnor and Cooperstock \cite{Bonnor}
and further discussed by Herrera and Varela \cite{Herrera1,Herrera2}.
We point out too, that the study of the mass in the RN spacetime
might help to find
the corresponding one in the Kerr spacetime. On the Kerr horizon mass
\cite{Kulkarni} is similarly characterized as in RN spacetime which we remark
in the next section.

To study the mass in RN spacetime we determine the
corresponding Schwarzschild spacetime, henceforth, the
Schwarzschild mass parameter, thus obtained, corresponds to the so called
geometrical mass in the RN spacetime. In order to obtain this correspondence
we study a class of solutions where we impose that the metric component
$g_{tt}$ is functionally related to the electrostatic potential $\phi$.
This technique of determining solutions of Einstein and Maxwell equations is
not new. It started with a remarkable paper by Weyl in 1917 \cite{Weyl},
where he found a class of electrostatic axially symmetric solutions
by imposing that $g_{tt}(\phi)$. Majumdar in 1947 \cite{Majumdar}
generalized this result to systems without spatial symmetry. The inclusion
of solutions with magnetostatic fields came through the works of
Papapetrou in 1947 \cite{Papapetrou} and Bonnor in 1954 \cite{Bonnor1}.
But in 1955 Ehlers \cite{Ehlers} gave a new approach to generate solutions
of Einstein and Maxwell equations by starting from given vacuum solutions.
Later, in the same vein, other solutions were found by Bonnor in 1961
\cite{Bonnor2} and Janis, Robinson and Winicour in 1967 \cite{Janis}.
However, all these solutions have the handicap of not switching back, in a
simple way, to its original vacuum solution. This difficulty was overcome
in 1976 by Teixeira, Wolk and Som (TWS). By using the operation of duality
rotation they were able to introduce simultaneously electrostatic and
magnetostatic fields, having the feature that by a proper choice of the
constants, the original vacuum solution emerged in a straightforward way.
This work generalizes the previous results \cite{Ehlers,Bonnor2,Janis}.
In 1977 Som, Santos and Teixeira \cite{SST} applied the TWS method to
obtain the RN solution from the Schwarzschild solution. This result was
reobtained recently \cite{FK}.

The plan of the paper is as follows. In section 2 we derive the method
developed by TWS \cite{TWS} in its inverse form, i.e.,
given a solution of the Einstein and Maxwell equations how to find
its corresponding vacuum
solution. Next, we apply the inverse method to obtain the corresponding
Schwarzschild spacetime to the RN spacetime. By doing this we deduce the
geometrical mass in the RN spacetime. In sections 3 and 4 we obtain the
active gravitational mass and mass function in the RN spacetime
and compare them to
the results of section 2. There is a discussion in the last section.

\section{The geometrical mass}
Here we present a method for deriving a class of vacuum solutions of
Einstein's equations out of a given solution of Einstein and Maxwell
equations. The class of solutions that we search for is when
we impose that the $g_{tt}$
metric component of the vacuum solution be functionally related
to the same component of the electrovacuum solution.
This method is the inverse of the TWS method \cite{TWS}. Then
we apply this method to deduce the Schwarzschild solution from the RN
solution.

\subsection{The inverse of the TWS method}
We consider the static line element corresponding to the electrovac
solution
\be
ds^2=e^{2\psi}dt^2-e^{-2\psi}h_{ij}dx^i dx^j,
\label{I-1}
\ee
where  $\psi$ and $h_{ij}$ are functions of the spatial components
$x^k$ (latin indices run from
1 to 3).  The Maxwell's equations in empty space read
\bea
{F^{\m\n}}_{;\m}=0,\label{I-2}\\
\e^{\m\n\rho\s}F_{\m\n;\rho}=0,
\label{I-3}
\eea
with $\e^{\m\n\rho\s}$ the totally antisymmetric tensor (greek indices run
from 0 to 3) with convention $\e^{0123}=1$. From (\ref{I-1}) and (\ref{I-3})
the electromagnetic
field tensor $F_{\m\n}$ can be written with the static non-null components
\be
F_{0i}=-\phi_{,i},
\label{I-4}
\ee
where $\phi$ is the electrostatic potential. Substituting (\ref{I-1})
and (\ref{I-4}) into (\ref{I-2}) we obtain
\be
(h^{1/2}e^{-2\psi}h^{ij}\phi_{,i})_{,j}=0,
\label{I-5}
\ee
with $h$ being the determinant of the spatial metric $h_{ij}$. Einstein
electrovac equations read
\be
G_{\m\n}=\kappa E_{\m\n}=\kappa\left(g^{\a\b}F_{\a\m}F_{\b\n}-
\frac{1}{4}g_{\m\n}F_{\a\b}F^{\a\b}\right),
\label{I-6}
\ee
where $G_{\m\n}$ is the Einstein tensor and $E_{\m\n}$ is the
electromagnetic
energy tensor. Substituting (\ref{I-1}) and (\ref{I-4}) into (\ref{I-6})
gives
\bea
e^{2\psi}h^{-1/2}(h^{1/2}h^{ij}\psi_{,i})_{,j}=h^{km}\phi_{,k}\phi_{,m},
\label{I-7}\\
H_{ij}+2\psi_{,i}\psi_{,j}=-2e^{-2\psi}\phi_{,i}\phi_{,j},
\label{I-8}
\eea
where $H_{ij}$ is the Ricci tensor in the 3 dimensional space.

If we consider now the static line element representing a solution of the
Einstein equations in the vacuum (as opposed to the electrovacuum):
\be
ds^2=e^{2V}dt^2-e^{-2V}h_{ij}dx^idx^j,
\label{I-9}
\ee
with $V$ a function of $x^k$, the corresponding Einstein equations read:
\bea
(h^{1/2}h^{ij}V_{,i})_{,j}=0,
\label{I-11}\\
H_{ij}+2V_{,i}V_{,j}=0.
\label{I-12}
\eea

We can then formulate the inverse of the TWS problem. Starting from a
static
solution $(\psi, h_{ij}, \phi)$ of Maxwell equations (\ref{I-5}) and
Einstein
electrovac equations (\ref{I-7}) and (\ref{I-8}), we want to obtain the
corresponding Einstein vacuum
solution $(V, h_{ij})$. We then search for a class of solutions where
$V$ is functionally related to $\psi$.

 From (\ref{I-5}) and (\ref{I-11}), we obtain
\be
aV_{,i}=e^{-2\psi}\phi_{,i},
\label{I-13}
\ee
where $a$ is an integration constant. Substituting (\ref{I-12}) into
(\ref{I-8}) and considering (\ref{I-13}), we get
\be
V_{,i}=\left(1+a^2e^{2\psi}\right)^{-1/2}\psi_{,i}.
\label{I-14}
\ee
By integration, (\ref{I-14}) leads to
\be
e^{2V}=e^{2\psi}\le[\frac {1+(1+a^2)^{1/2}}{1+(1+a^2e^{2\psi})^{1/2}}\ri]^2,
\label{I-15}
\ee
where we have chosen the integration constant such that when $a=0$, the
electric field is zero and the electrovac solution (\ref{I-1}) reduces to
the vacuum solution (\ref{I-9}).
Introducing (\ref{I-13}) into (\ref{I-7}) with (\ref{I-14}) yields
\be
h^{-1/2}(h^{1/2}h^{ij}\psi_{,i})_{,j}=
\frac {a^2e^{2\psi}}{1+a^2e^{2\psi}}h^{km}\psi_{,k}\psi_{,m},
\label{I-16}
\ee
which means that the constant $a$ is determined by the parameters involved
in $\psi$ and $h_{ij}$.

\subsection{Schwarzschild solution from a RN solution}
We now apply the preceding method to determine the Schwarzschild
solution associated with a RN metric. First, substituting the RN line
element,
\be
ds^2=\le(1-\frac {2M}{r}+\frac {Q^2}{r^2}\ri)dt^2-\le(1-\frac {2M}{r}+
\frac {Q^2}{r^2}\ri)^{-1}dr^2-r^2d\O^2,
\label{I-17}
\ee
into (\ref{I-16}) we find for $a$
\be
aM_S=Q,
\label{I-18}
\ee
where
\be
M_S=(M^2-Q^2)^{1/2}.
\label{I-19}
\ee
We see from (\ref{I-18}) that when $a=0$ then $Q=0$, reducing the RN
spacetime (\ref{I-17}) to the vacuum Schwarzschild spacetime.
The term $e^{2\psi}$ in (\ref{I-1}) can be expressed from (\ref{I-17}),
with the aid of (\ref{I-19}), as
\be
e^{2\psi}=\frac {(r-M_S-M)(r+M_S-M)}{r^2}.
\label{I-21}
\ee
Substituting (\ref{I-18}) and (\ref{I-21}) into (\ref{I-15}) we obtain
\be
e^{2V}=1-\frac{2M_S}{r+M_S-M}.
\label{I-22}
\ee
Considering (\ref{I-22}), the vacuum solution (\ref{I-9}) becomes
\be
ds^2=\le(1-\frac {2M_S}{R}\ri)dt^2-\le(1-\frac {2M_S}{R}\ri)^{-1} dR^2-
R^2d\O^2,
\label{I-23}
\ee
where $R$ is given by
\be
R=r+M_S-M.
\label{I-24}
\ee
Hence, the corresponding vacuum solution to the RN spacetime is the
Schwarzschild spacetime (\ref{I-23}) with geometrical mass
$M_S=(M^2-Q^2)^{1/2}$ and with a scale shift (\ref{I-24}) in the radial
coordinate $R$.

The event horizon $r_{eh}$ and the Cauchy horizon $r_{Ch}$ for the RN
spacetime (\ref{I-17}) are then
\be
r_{eh}=M+M_S, \;\; r_{Ch}=M-M_S,
\label{I-24a}
\ee
while for the corresponding Schwarzschild spacetime (\ref{I-23}) these
two surfaces represent, respectively, the event horizon $R_{eh}$ and
singularity $R_0$,
\be
R_{eh}=r_{eh}+M_S-M=2M_S, \;\; R_0=r_{Ch}+M_S-M=0.
\label{I-24b}
\ee

We observe that the Kerr spacetime effective mass $M_{eff}$, as obtained in
\cite{Kulkarni}, on its event horizon is
\be
(M_{eff})_{eh}=(M_K^2-a^2)^{1/2},
\label{I-24c}
\ee
where $M_K$ is the Kerr mass and $a$ its specific angular momentum.
 From (\ref{I-19}) and (\ref{I-24c}) we have that the effective Kerr mass
resembles the geometric RN geometrical mass at least on the event
horizon, thus suggesting some similarity between both spacetimes.

\section{The active gravitational mass}
The active gravitational mass density $\m$, given by Tolman \cite{Tolman}
and Whittaker \cite{Whittaker}, is for the field equations (\ref{I-6})
\be
\m={E^0}_0-{E^i}_i,
\label{II-1}
\ee
and the total active gravitational mass within a volume $V$, coming out
from the work of Whittaker \cite{Whittaker} reads
\be
M_a=\int_V\m(-g)^{1/2}dx^1dx^2dx^3,
\label{II-2}
\ee
where $g$ is the four dimensional determinant of the metric. Applying
(\ref{II-2}) to the RN spacetime (\ref{I-17}) we find
\be
M_a(\infty)-M_a(r)=\int_{r}^\infty \frac {Q^2}{r^2}dr,
\label{II-2a}
\ee
which, assuming that $M_a(\infty)=M$, leads to
\be
M_a(r)=M-\frac{Q^2}{r}.
\label{II-3}
\ee
The active mass is  positive or null for $r\geq r_0=Q^2/M$ with
$r_{Ch}\leq r_0\leq r_{eh}$. It can take negative values for $r<r_0$ and
this possibility has been
discussed in a number of papers \cite{Bonnor,Cruz,CG,Cooperstock}.
The existence of negative energies comes from the fact \cite{Papapetrou}
that since the electrostatic energy of a point charge is infinite, then to
have its total energy finite there must be an infinite amount of
negative energy at its center of symmetry.
The active gravitational mass (\ref{II-3}) for the event horizon and
Cauchy horizon are, respectively,
\be
M_a(r_{eh})=M_S, \;\; M_a(r_{Ch})=-M_S.
\label{II-4}
\ee
The spacetime external to $r=r_{eh}$ is very similar to the
Schwarzschild spacetime external to the surface $r=2M_S$ (see discussion
in \cite{Chandra} p. 209). Hence the surface $r=r_{eh}$ is an event
horizon
in the same sense that $r=2M_S$ in Schwarzschild spacetime. However, the
spacetime internal to $r=r_{eh}$ has a completely different structure as
compared to the Schwarzschild spacetime (refer again to \cite{Chandra}).
It is interesting that the active gravitational mass (\ref{II-3})
measures in RN spacetime the corresponding Schwarzschild mass at
$r=r_{eh}$.

The electric field $E$ due to (\ref{I-17}) is
\be
E=\frac{Q}{r^2},
\label{II-5}
\ee
and from (\ref{I-5}) the corresponding non-relativistic Maxwell energy,
$M_E(r,\infty)$,
entrapped outside the spherical surface of radius $r$ is given by
\be
M_E(r,\infty)=\frac{1}{2}\int^{\infty}_rE^2r^2dr.
\label{II-6}
\ee
Using (\ref{II-5}) and (\ref{II-6}), the active gravitational mass
(\ref{II-3}) can be rewritten
\be
M_a(r)=M-2M_E(r,\infty).
\label{II-8}
\ee
 From (\ref{II-8}) we have that $M_a(r)$ is the total active gravitational
mass minus twice the mass equivalent to the non-relativistic energy stored
by the electric field
$E$ outside the spherical surface of radius $r$.

Calculating the circular geodesics in the equatorial plane
and the radial geodesics \cite{Gron} of RN spacetime for a
chargeless particle we
obtain, respectively, from (\ref{I-17}) and (\ref{II-3}),
\be
\frac{d^2r}{d\t^2}=-\frac{1}{r^2}\left(M-\frac{Q^2}{r}\right)=
-\frac{M_a(r)}{r^2}, \;\; \left(\frac{d\phi}{d\tau}\right)^2=
\frac{1}{r^3}\left(M-\frac{Q^2}{r}\right)=\frac{M_a(r)}{r^3},
\label{II-9}
\ee
where $\tau$ is the proper time. The motion of a chargeless particle will
then be affected by the charge of the black hole eventually producing
repulsive
forces for sufficiently small values of $r$. The active
gravitational mass casts locally the equations of motion in a Newtonian
like form.

\section{The mass function}
There exists a further different way to define the gravitational mass
of a system.
If we match a spherical distribution of matter to the exterior
Schwarzschild
spacetime we obtain at the surface of discontinuity the Schwarzschild
mass being equal to $rR^{\phi}_{\th\phi\th}/2$ where
$R^{\phi}_{\theta\phi\theta}$ is a Riemann tensor component.
The interpretation of this
mass as the total mass inside the sphere suggests that the total mass
entrapped inside a sphere of radius $r$, called the mass function,
may be defined by
\be
M_f(r)=\frac{1}{2}r{R^{\phi}}_{\th\phi\th}.
\label{III-1}
\ee
This definition has been first considered by Lema\^{\i}tre \cite{Lemaitre}
and since currently used in gravitational collapse
\cite{Misner,May,Wilson,Glass,Bruenn,Burrows,Adams}.
For RN spacetime (\ref{I-17}) we have for (\ref{III-1})
\be
M_f(r)=M-\frac{Q^2}{2r},
\label{III-2}
\ee
where
\be
M=\frac{1}{2}r{C^{\phi}}_{\th\phi\th},
\label{III-3}
\ee
${C^{\phi}}_{\theta\phi\theta}$ being a Weyl tensor component.
$M$ is called the pure gravitational mass, since it arises only from the
Weyl tensor. The mass function is always positive or null for $r\geq
r_1=Q^2/2M$ with $r_1=r_0/2$ inside the event horizon. Negative values for
$M_f(r)$ for $r<r_1$ are considered in \cite{CMacV}. Here applies the same
remark about (\ref{II-8}) concerning negative energies.
The mass function (\ref{III-2}) at the event horizon and Cauchy horizon
(\ref{I-24a}) are, respectively,
\be
M_f(r_{eh})=\frac{1}{2}(M+M_S), \;\; M_f(r_{Ch})=\frac{1}{2}(M-M_S).
\label{III-4}
\ee

Considering (\ref{II-5}) and (\ref{II-6}) we have
\be
M_f(r)=M-M_E(r,\infty).
\label{III-6}
\ee
The fact that in (\ref{II-8}) $M_E(r,\infty)$ is multiplied by 2 reflects
that in (\ref{II-2}) not only pure gravitational energy produced by the
charge is considered like in (\ref{III-1}).

\section{Discussion}
We have presented the TWS method \cite{TWS} in its inverse form and
applied it to obtain the vacuum Schwarzschild solution from the electrovacuum
RN solution. The vacuum metric thus derived has a geometrical mass $M_S$
(\ref{I-18}). We observe that the structure of $M_S$ is similar to the
Kerr effective mass $M_{eff}$ (\ref{I-24c}) obtained in \cite{Kulkarni}
on its event horizon.
Then we calculate for RN spacetime its active gravitational
mass (\ref{II-3}) and mass function (\ref{III-2}). Both results are equal
to $M$ asymptotically but, in general, differ. This illustrates
the ambiguity in the localization of energy. We compare $M_a(r)$ and
$M_f(r)$ with $M_S$ at the horizons, respectively (\ref{II-4}) and
(\ref{III-4}), and obtain that $M_a(r_{eh})=M_S$. This result is not so
surprising because of the following
reasons. The structure of the
event horizon for RN spacetime and Schwarzshild spacetime are similar
\cite{Chandra}. The equations of motion
(\ref{II-9}) are cast in a locally Newtonian like form with the aid of the
active gravitational mass. Furthermore,
by analysing the energy content of a slowly collapsing gravitating sphere,
Herrera and Santos \cite{Herrera}
conclude that the active gravitating mass grasps better the physical
content of matter than the mass function. Hence we can say that our
results give further support the physical meaningfulness of $M_a(r)$.

\section*{Acknowledgment}
We thank Fred Cooperstock for kindly helping to improve the presentation
of the paper.
NOS gratefully acknowledges financial assistance from CNPq, Brazil.

\end{document}